\documentclass[conference]{IEEEtran}
\IEEEoverridecommandlockouts
\usepackage{subfigure}
\usepackage[]{graphicx, amsfonts, amsmath, amssymb, amstext, latexsym, float, color}

\newcommand{\be}{\begin{equation}}
\newcommand{\ee}{\end{equation}}
\newcommand{\ber}{\begin{eqnarray}}
\newcommand{\eer}{\end{eqnarray}}

\pdfoutput=1
\newtheorem{theorem}{Theorem}

\newtheorem{lemma}[theorem]{Lemma}

\newtheorem{proposition}[theorem]{Proposition}
\newtheorem{remark}[theorem]{Remark}

\newenvironment{proof}[1][Proof]{\noindent\textbf{#1.} }{\ \rule{0.5em}{0.5em}}

\usepackage{stmaryrd} 
\usepackage{array} 
\usepackage{multirow} 
\usepackage[disable]{todonotes} 

\DeclareMathOperator{\sinc}{sinc}
\DeclareMathOperator{\Si}{Si}
\DeclareMathOperator{\argmax}{argmax}
\DeclareMathOperator{\emean}{\mbox{\textbf{E}}}
\DeclareMathOperator{\MSE}{\text{MSE}}
\DeclareMathOperator{\NRMSE}{\text{NRMSE}}
\newcommand{\etal}{\textit{et al.}\ }
\newcommand{\est}[1]{\widehat{#1}}
\newcommand{\real}[1]{\text{Re}\left[#1\right]}

\begin{document}

\title{Fundamental limit of resolving two point sources limited by an arbitrary point spread function}

\author{
	\IEEEauthorblockN{Ronan Kerviche}
	\IEEEauthorblockA{University of Arizona\\
	Email: rkerviche@optics.arizona.edu} 
	\and
	\IEEEauthorblockN{Saikat Guha}
	\IEEEauthorblockA{Raytheon BBN Technologies\\
	Email: saikat.guha@raytheon.com}
	\and
	\IEEEauthorblockN{Amit Ashok}
	\IEEEauthorblockA{University of Arizona\\
	Email: ashoka@optics.arizona.edu}

	\thanks{This work was supported by the DARPA REVEAL program under contract number HR0011-16-C-0026. While preparing this paper we became aware of related work~\cite{Rehacek16}, which has some overlap with results presented in this paper. All the detailed proofs are relegated to an Appendix at the end of the paper.}
}

\maketitle
\begin{abstract}
Estimating the angular separation between two incoherently radiating monochromatic point sources is a canonical toy problem to quantify spatial resolution in imaging. In recent work, Tsang {\em et al.} showed, using a Fisher Information analysis, that Rayleigh's resolution limit is just an artifact of the conventional wisdom of intensity measurement in the image plane. They showed that the optimal sensitivity of estimating the angle is only a function of the total photons collected during the camera's integration time but entirely independent of the angular separation itself no matter how small it is, and found the information-optimal mode basis, intensity detection in which achieves the aforesaid performance. We extend the above analysis, which was done for a Gaussian point spread function (PSF) to a hard-aperture pupil proving the information optimality of image-plane sinc-Bessel modes, and generalize the result further to an arbitrary PSF. We obtain new counterintuitive insights on energy vs. information content in spatial modes, and extend the Fisher Information analysis to exact calculations of minimum mean squared error, both for Gaussian and hard aperture pupils. 
\end{abstract}

\vspace{-2mm}
\section{Introduction and Background}\label{sec:intro}

Consider estimating the angular separation $2\theta$ between two incoherently-radiating $\lambda$-wavelength quasi monochromatic point sources in the far field that are symmetrically disposed about the line of sight. The aperture of the camera has diameter $D$, and during the integration time the total mean photon number collected is denoted $N$. A conventional camera uses a lens in the plane of the aperture pupil to focus the image in an {\em image plane}, and detects the image-plane intensity pattern using a detector pixel array. The field amplitude in the image plane is an aperture-blurred version of the true object profile, i.e., a scaled version of convolution of the object-plane field (two independent delta functions for the above problem) with the amplitude spread function (ASF) $A(x)$ of the camera's aperture. It is well known that no matter what $\theta$ is, the minimum mean squared error (MMSE) of estimating $\theta$ can be made arbitrarily small by using a long enough exposure, i.e., by taking $N \to \infty$. Rayleigh showed that, for a conventional camera that measures the intensity in the image plane, even if that intensity measurement is done with infinitely many infinitesimally-tiny shot-noise-limited detector pixels, when $\theta$ decreases below $\sim \lambda/D$~\cite{Rayleigh}, the mean squared error (MSE) of estimating $\theta$ drastically degrades (increases) for the integration time (hence $N$) held fixed. In a recent breakthrough result, Tsang {\em et al.} showed that, assuming a Gaussian ASF, intensity measurement in the infinite Hermite-Gauss (HG) basis in the image-plane coordinates attains a Fisher Information ${\cal{I}}_{\rm HG}(\theta)$ that is independent of $\theta$ no matter how small is $\theta$, and equals the high-$\theta$ MSE attained by conventional image-plane direct detection~\cite{MTsang161}. They also showed that the quantum Fisher Information (QFI) ${\cal{I}}_{\rm Q}(\theta)$ for estimating $\theta$---which makes no assumptions on how the optical field collected by the aperture gets pre-processed and detected---equals ${\cal{I}}_{\rm HG}(\theta)$, establishing that a linear spatial mode sorting prior to detection, which separates mutually-orthogonal HG modes and detects each with individual detector pixels, is an optimal detector for this problem. This showed that Rayleigh's criterion is an artifact of the conventional philosophy of intensity measurement in the image plane. There is rich information content in the phase of the image to extract, which optimally using a shot-noise-limited intensity measurement, one must use a non-trivial spatial-mode transformation to the aperture field prior to detection to manipulate the post-detection shot noise so as to maximize the information content about $\theta$ in the noisy detection outcomes.

This result opens up a variety of interesting questions, some important ones being: (a) what is the information-optimal mode basis for the two-point-source problem with a hard-aperture pupil (sinc ASF) and for other general ASFs, (b) what is the right minimal set of modes that carry almost all the relevant information about a passive imaging problem, (c) what is the actual advantage in MSE (Fisher Information provides a lower bound on the MMSE, via the Cramer Rao lower bound (CRLB), which is not always achievable), and (d) how does this theory generalize to more complex imaging problems, and to broadband light.

In this paper, we focus on the two-point-source problem described above. Our contributions are summarized below:

\noindent {\bf 1.} With a rectangular hard aperture, i.e. sinc ASF, we show that measuring in the sinc-Bessel (SB) mode basis achieves the QFI and the Fisher Information is independent of $\theta$, analogous to measuring in the HG mode basis with a Gaussian aperture.

\noindent {\bf 2.}  We evaluate the exact MMSE of estimating $\theta$ with the optimal mode basis and compare with the CRLB.

\noindent {\bf 3.} We illustrate a counterintuitive distribution of energy vs. information in the individual modes of the optimal mode basis, which provide insights on efficient measurement design. In particular we find that if one were to extract and detect a single mode and its orthogonal complement (e.g., the binary SPADE of~\cite{MTsang161}), the $1^{\rm st}$ mode is optimal rather than the $0^{\rm th}$ mode, in the deep sub-Rayleigh limit. We discuss information loss due to leaky mode separation. 

\noindent {\bf 4.}  We provide the optimal binary SPADE measurement for a general ASF which attains the QFI in the low $\theta$ limit, and new insights into constructing information optimal mode bases.

\section{Optimal modes for hard aperture}

	
	
Let $A(x)$ be the (generally complex-valued) energy-normalized Amplitude Spread Function (ASF) of the aperture, i.e., $\int_{-\infty}^{+\infty} |A(x)|^2\ dx = 1$. The image plane field is an incoherent sum of two symmetrically-shifted copies of the ASF at $\pm\theta$, each of which is perfectly self coherent. If one projects the image plane field onto the complex-valued spatial mode $f(x)$, $\int_{-\infty}^{+\infty} |f(x)|^2\ dx = 1$, the fraction of the intensity in the image-plane field that appears in the $f(x)$ mode is given by:
	\begin{align}
		m_f(\theta) &= \frac{1}{2}\left|\int_{-\infty}^{+\infty} \overline{f(x)} A(x+\theta)\ dx\right|^2 \nonumber\\
		&\quad + \frac{1}{2}\left|\int_{-\infty}^{+\infty} \overline{f(x)} A(x-\theta)\ dx\right|^2, \label{twoPointsMeasurementDefinition}
	\end{align}
	\todo{Describe that this is the fractional power}
where $\overline{f(x)}$ denotes complex conjugate. We will call $m_f(\theta)$ the {\em measurement function} for mode $f$. In~\cite{MTsang161}, the authors showed that if $A(x)$ is Gaussian, then projecting the image-plane field simultaneously onto the infinite Hermite Gauss (HG) orthonormal mode basis functions $f_q(x)$, $q=0, 1, \ldots$, attains a vector of measurements whose Fisher Information content on $\theta$ is quantum optimal, and independent of $\theta$. We will develop a similar strategy but for the practically relevant case of a space-limited (hard) aperture. This produces a cardinal sine ($\sinc$) ASF, i.e., $A(x) = {\sinc}(x)$.

In order to match the energy distribution of the PSF, we will choose an orthonormal basis for which the ASF is the first basis function. This is the case of the Spherical Bessel Functions of the First Kind, $M_q(x) = \sqrt{1+2q}\,j_q(\pi x), q\in\mathbb{N}$, which are all either even or odd. Note that $j_0(\pi x) = \sinc(x)$. 
	It is simple to verify that $\int M_q(x) \sinc(x'-x)\ dx = M_q(x')$. We will introduce a spatial scale factor $\sigma$ of dimensions of length in the image plane coordinate ($x$) to capture the actual `length' of the ASF in the image plane. It will depend upon the diameter of the aperture and the focal length of the imaging system. The fraction of the total energy collected in the $q^{\rm th}$ mode is given by:
	\begin{align}
		m_q(\theta) &= M_q\left(\frac{\theta}{\sigma}\right)^2. \label{sincMeasurements}
	\end{align}

	From \cite{Abramowitz64}, equation 1.10.50 :
	\begin{align}
		\forall x\in\mathbb{R},\ \sum_{q=0}^\infty (1+2q) j_q(x)^2 = 1 \Rightarrow \sum_{q=0}^\infty m_q(x) = 1, \label{sincEnergyEfficiency}
	\end{align}
	shows that the sinc Bessel (SB) modes capture all the energy in the image-plane field. With $N$ being the total mean photon number collected over the camera's integration time, the number of photons in the $q^{\rm th}$ SB mode is $N\,m_q(\theta)$.

	Each of separated SB modes is detected using a shot-noise-limited detector. The total number of orthogonal temporal modes in the collected field $M \approx T(\Delta \nu)$, where $T$ is the integration time and $\Delta \nu$ is the bandwidth of the light around its center frequency. The number of photons per temporal mode $N_0 \ll 1$ at optical frequencies. $N = MN_0$. It is simple to show that with $N_0 \ll 1$ and $M \gg 1$, and with the photon statistics in the individual modes being distributed with the thermal (geometric, or Bose Einstein) distribution, that the total number of photons in the $q^{\rm th}$ spatial mode is Poisson distributed with mean $N\,m_q(\theta)$. In order to calculate the Fisher Information of $\theta$ in the $q^{\rm th}$ SB mode, we will rely on the following result:
	\begin{lemma}[Fisher Information For A Poisson Corrupted Process]
		Let $f$ be a $\mathcal{C}_1$ function with values in $\mathbb{R}^+$ mapping the variable of interest $\theta$ to a measurement $Y \sim \mathcal{P}\left(y|f(\theta)\right)$ where $\mathcal{P}$ is the Poisson distribution. Then the Fisher Information associated to the process can be written as:
		\begin{align}
			\mathcal{I}(\theta) = \frac{f'(\theta)^2}{f(\theta)}.
		\end{align}
	\end{lemma}

	This expression is similar to that obtained for a Gaussian-corrupted measurement process, $\mathcal{I}(\theta) = f'(\theta)^2/\eta^2$, where $f'(\theta)$ is the sensitivity and $\eta^2$ is the noise variance. The Fisher Information in the $q^{\rm th}$ SB mode evaluates to: 
	\begin{align}
		\mathcal{I}_q(\theta) &= N \frac{m_q'(\theta)^2}{m_q(\theta)} \\
		&= \frac{4\pi^2N}{\sigma^2} (1+2q) \left(\frac{q\sigma}{\pi\theta}j_q\left(\frac{\pi\theta}{\sigma}\right) - j_{q+1}\left(\frac{\pi\theta}{\sigma}\right)\right)^2. \label{sincFisherInformation}
	\end{align}

	\begin{lemma}[Series of Spherical Bessel Function Of The First Kind]\label{lem:sumSB_FI}
		We have the following result on a series of Spherical Bessel Function of the First Kind on any finite interval $I$ of $\mathbb{R}$ and containing $0$ (see Appendix for proof):
		\todo{Replace $I$ by $E$ to avoid confusion with the intensity?}
		\begin{align}
			\forall x\in I,\ \sum_{q=0}^\infty &(1+2q)\frac{q^2}{x^2}j_q(x)^2 + (2+4q)\frac{q}{x}j_q(x)j_{q+1}(x) \nonumber \\
			&+ (1+2q)j_{q+1}(x)^2 = \frac{1}{3}.
		\end{align}
	\end{lemma}

	The measurement outputs on any orthogonal mode set are statistically independent Poisson random variables. So, the total Fisher Information for the vector-parametrized estimator is equal to the sum of the individual Fisher Informations from each mode. Using Lemma~\ref{lem:sumSB_FI}, we deduce that measuring all the SB modes leads to a $\theta$-independent Fisher Information, \textit{i.e.},
	\begin{align}
		\forall \theta\in I,\ \mathcal{I}(\theta) = \sum_{q=0}^\infty \mathcal{I}_q(\theta) = \frac{4\pi^2 N}{3\sigma^2}. \label{sincFisherInformationBound}
	\end{align}
In~\cite{MTsang161}, it was shown that $\mathcal{I}(\theta)$ in~\eqref{sincFisherInformationBound} is the QFI for estimating $\theta$ with a hard aperture. Hence, we now have a proof that a specific SB mode sorting prior to direct detection achieves the QFI, in the sense that the classical Fisher Information of the SB-mode measurement exactly matches the QFI.
	\todo{Missing : proof that is actually the best achievable information?}

\section{Energy vs. Information Content in Modes}


	
With the Gaussian ASF $A(x) = (2\pi\sigma^2)^{-\frac{1}{4}}\exp(-x^2/4\sigma^2)$, the measurement function and Fisher information in the $q^{\rm th}$ image-plane HG mode, $q\in\mathbb{N}$, are respectively given by~\cite{MTsang161}:
	\begin{eqnarray}
		m_q(\theta) &=& \frac{1}{q!}\left(\frac{\theta^2}{4\sigma^2}\right)^q e^{-{\theta^2}/{4\sigma^2}}, \,{\text{and}} \label{gaussianMeasurements} \\
		\mathcal{I}_q(\theta) &=& \frac{N}{\sigma^2 q!} \left(q - \frac{\theta^2}{4\sigma^2}\right)^2 \left(\frac{\theta^2}{4\sigma^2}\right)^{q-1}e^{-{\theta^2}/{4\sigma^2}}. \label{gaussianFisherInformation}
	\end{eqnarray}
The total Fisher information from measuring all the (infinitely many) HG modes equals the QFI bound for any $\theta$, i.e.~\cite{MTsang161},
	\begin{align}
		{\mathcal I}(\theta) = \sum_{q=0}^\infty \mathcal{I}_q(\theta) &= \frac{N}{\sigma^2}.
		\label{eq:sumHGinformation_Gaussian}
	\end{align}
The Fisher Information attained by infinite-spatial-resolution image-plane direct detection is given by,
	\begin{align}
		\mathcal{I}_{\text{Direct}} = \int_{-\infty}^{\infty} \frac{I'(x, \theta)^2}{I(x, \theta)}\ dx, \mbox{  with}\ I'(x, \theta) = \frac{\partial I(x, \theta)}{\partial \theta}, \label{directFisherInformation}
	\end{align}
	where $I(x,\theta) = |A(x,\theta)|^2$ is the normalized spatial distribution of energy in the image plane, also the probability density function of measuring a photon at spatial position $x$, conditioned on $\theta$. $\mathcal{I}_{\text{Direct}}$ approaches the QFI in~\eqref{eq:sumHGinformation_Gaussian} for $\theta \to \infty$, but goes to zero as $\theta \to 0$. So, the information advantage of the HG mode measurement over image-plane direct detection is maximum at small $\theta$ (sub Rayleigh regime)~\cite{MTsang161}.

Comparing~\eqref{gaussianMeasurements} and~\eqref{gaussianFisherInformation} with ~\eqref{sincMeasurements} and~\eqref{sincFisherInformation}, we see several analogous trends. First, measuring all the modes (HG or SB, respectively) in either case captures all the image-plane energy for any $\theta$. Hence, any spatial mode orthogonal to the span of the respective mode sets would neither have any energy nor any information content. Further, in both cases the $q=0$ mode captures all of the energy at $\theta = 0$. Finally, as noted in~\cite{MTsang162}, only the $q=1$ HG mode contributes to the total Fisher information ${\mathcal I}(\theta)$ at low $\theta$ for the Gaussian aperture case. With a hard aperture, the $q=1$ SB mode has that exact same property.

In what follows, we will consider information contributions from individual modes and in sets of modes (without resolving modes in the set). We first consider the following lemma:
	\begin{lemma}[Fisher Information Inequality On Aggregated Measurements]
		Consider measurement functions $\left\{m_q(\theta)\right\}$ corresponding to an orthonormal family of modes. Let us say we make an aggregated measurement where we project the image plane field on to a collection of modes $S$, i.e., an effective measurement function $m_S(\theta) = \sum_{q \in S} m_q(\theta)$. This measurement cannot give us more information than the sum of the information in the individual modes in the set, \textit{i.e.},
		\begin{align}
			\theta\in\mathbb{R},\ \mathcal{I}_S(\theta) = \frac{m_S'(\theta)^2}{m_S(\theta)} \le \sum_{q \in S} \mathcal{I}_q(\theta) = \sum_{q \in S} \frac{m_q'(\theta)^2}{m_q(\theta)}. 
		\end{align}
	\end{lemma}

We now want to find a single mode $g$ whose information content does not go to zero (i.e., goes to a constant $c$) as $\theta \to 0$. To see the requirement on $g$, let us consider the following:
	\begin{proposition}[Insensitivity Property]
		Given any properly normalized and continuous mode $g(x)$ over $\mathbb{R}^+$ and any ASF $A(x)$, the first order derivative $m_g'(\theta)$ of the associated measurement function $m_g(\theta)$ goes to $0$ as $\theta \to 0^+$.
	\end{proposition}

	Hence, in order for the information content in mode $g$, $I_g(\theta) = m_g^\prime(\theta)^2/m_g(\theta)$ to go to a constant as $\theta \to 0^+$, the following equivalence relation must be satisfied:
	\begin{align}
		\exists c>0, {\text{s.t.}}\,\, c\, m_g(\theta) \underset{\theta\rightarrow 0^+}{\sim} m_g'(\theta)^2.
		\label{eq:equivalence}
	\end{align}
For the above to hold, it is necessary (but not sufficient in general) for $m_g \to 0$ as $\theta \to 0^+$. Thus, $m_g(\theta)$ should neither have sensitivity nor should it capture any energy at $\theta = 0$. Yet, if~\eqref{eq:equivalence} is satisfied, measuring $g$ will produce non zero information for $\theta \to 0^+$. Note that the insensitivity property applies to $I^\prime(x,\theta)$ in~\eqref{directFisherInformation} for the direct measurement as well: it equals $0$ regardless of the ASF. So an image-plane direct measurement provides no information about $\theta$ when $\theta \to 0^+$.

A binary SPADE (Bin-SPADE) receiver measures a single mode $g(x)$ and its orthogonal component (i.e., the leftover energy) leading to a simple implementation~\cite{MTsang161}. For both Gaussian and hard apertures, BinSPADE receivers constructed for $g(x)$ being the respective $q=0$ mode (the ASF mode) and one with the $q=1$ mode ($q=1$ HG or SB mode, respectively) attain a non-zero information at $\theta \to 0^+$, which equals the respective QFI limit. We will refer to these two receivers as 0-BinSPADE and 1-BinSPADE, respectively. The Fisher Information for these two measurements for the Gaussian ASF are given by:
	\begin{align}
		\mathcal{I}_{\text{0-BinSPADE}}(\theta) &= \frac{N}{\sigma^2} \frac{\frac{\theta^2}{4\sigma^2}}{\exp\left(\frac{\theta^2}{4\sigma^2}\right) - 1}, \,{\text{and}} \\
		\mathcal{I}_{\text{1-BinSPADE}}(\theta) &= \frac{N}{\sigma^2} \frac{\left(1 - \frac{\theta^2}{4\sigma^2}\right)^2}{\exp\left(\frac{\theta^2}{4\sigma^2}\right) - \frac{\theta^2}{4\sigma^2}},
	\end{align}
whereas, for sinc ASF (and SB modes), the respective Fisher Informations are given by:
	\begin{align}
		\mathcal{I}_{\text{0-BinSPADE}}(\theta) &= \frac{4N}{\sigma^2} \frac{\left(\frac{\sigma}{\theta}\cos\left(\frac{\pi\theta}{\sigma}\right) - \frac{\sigma^2}{\pi\theta^2} \sin\left(\frac{\pi\theta}{\sigma}\right)\right)^2}{1 - j_0\left(\frac{\pi\theta}{\sigma}\right)^2}, \\
		\mathcal{I}_{\text{1-BinSPADE}}(\theta) &= \frac{12N}{\sigma^2} \frac{\sigma^2}{\theta^2} \frac{\left[\sin\left(\frac{\pi\theta}{\sigma}\right) - 2 j_1\left(\frac{\pi\theta}{\sigma}\right)\right]^2}{1 - 3 j_1\left(\frac{\pi\theta}{\sigma}\right)^2}.
	\end{align}

\begin{figure}
		\begin{center}
			\includegraphics[width=8.0cm]{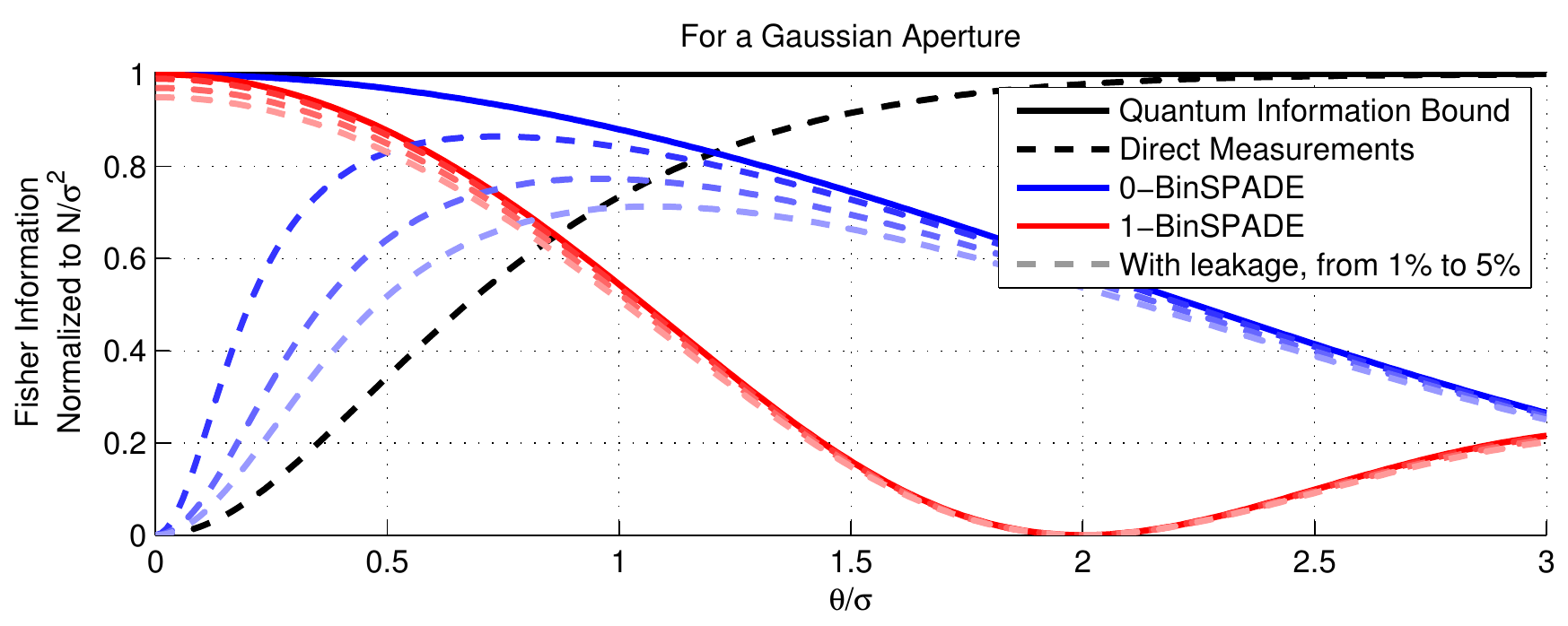}
			\includegraphics[width=8.0cm]{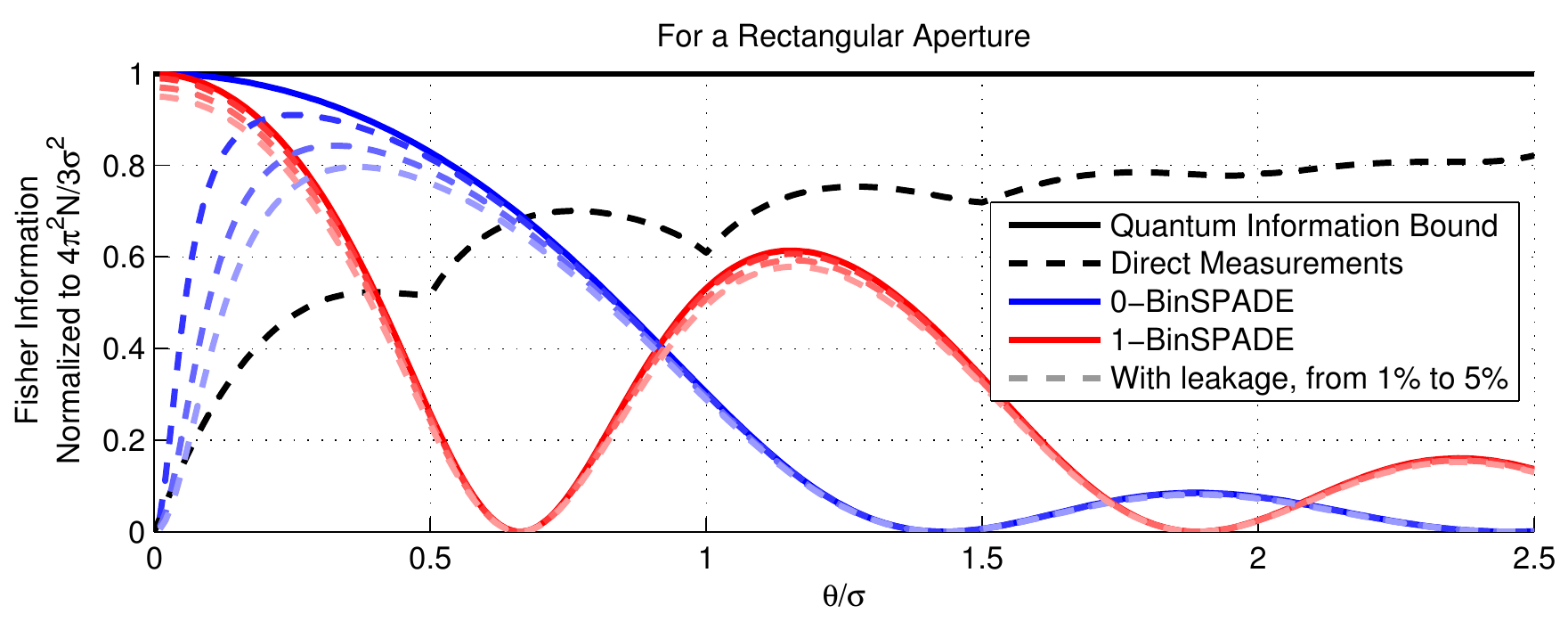}
		\end{center}
		\vspace{-6mm}
		\caption{Performance comparison of Direct Measurements versus 0 and 1-BinSPADE receivers for Gaussian (\textit{top}) and hard (\textit{bottom}) apertures. The y-axes are normalized to the respective Quantum Fisher Information bounds.}
		\vspace{-4mm}
		\label{FisherInformationPlots}
	\end{figure}

Even though both BinSPADE receivers attain the QFI for $\theta \to 0$, the former significantly outperforms the latter for higher $\theta$ (see Fig.~\ref{FisherInformationPlots}). However 1-BinSPADE is much more robust to imperfect implementation. Let us say $\epsilon > 0$ is a leakage parameter such that the BinSPADE receiver measures $(1-\epsilon)m_g(\theta)$ and its orthogonal complement. Then, absolutely any $\epsilon>0$ results in $\mathcal{I}_{\text{0-BinSPADE}}(\theta)$ to collapse to $0$ at $\theta = 0$. On the contrary, the performance of 1-BinSPADE at small $\theta$ remains fairly stable and leakage tolerant, with the loss in Fisher information being proportional to the energy loss in the detected mode, thus continuing to satisfy the condition stated above and hence retaining a non-zero information at $\theta \to 0^+$. 

\section{Mean Squared Error (MSE) Analysis}


Let us recall the following inequality between the Mean Square Error ($\MSE_{\est{\theta}}$) of an estimator $\est{\theta}$, its bias $B_{\est{\theta}}$, and the Fisher Information ${\mathcal I}(\theta)$:
	\begin{align}
		\MSE_{\est{\theta}}(\theta)\ge \frac{(1-B_{\est{\theta}}'(\theta))^2}{\mathcal{I}(\theta)} + B_{\est{\theta}}(\theta)^2,
	\end{align}
where $B_{\est{\theta}}'(\theta)$ is the first derivative of the bias function with respect to $\theta$. For comparing the MSE attained by mode-sorting receivers and image-plane detection, we will use the Normalized Root Mean Square Error, $\NRMSE_{\est{\theta}} = \sqrt{\MSE_{\est{\theta}}(\theta)}/\theta$ and the adapted Cramer-Rao bound defined as $1/\theta\sqrt{\mathcal{I}(\theta)}$~\cite{Rao45, Cramer46}.

For a set of independent and identically distributed (i.i.d.) measurements $\left\{y_q\right\}, q=0, \ldots, Q$ drawn from the conditional distribution $\mathcal{P}(y_q|\theta)$ (as will be the case when a set of orthogonal image-plane spatial modes are detected simultaneously), the Maximum Likelihood Estimator (MLE) is given by:
	\begin{align}
		\est{\theta}(y_0, \ldots, y_Q) = \argmax\left\{\prod_{q=0}^Q \mathcal{P}(y_q|\theta) \right\}.
	\end{align}

Assuming the prior knowledge of the total photon number $N$ collected during the integration time, the MLE for a measurement that just measures the $q(x)$ mode, is given by:
	\begin{align}
		\est{\theta}(y_q) = m_q^{-1}\left({y_q}/{N}\right).
	\end{align}
For the $q$-BinSPADE receiver (i.e., one that measures mode $q$ and its orthogonal complement), the total photon number can be inferred from the sum count over the two measurements: 
	\begin{align}
		\est{\theta}(y_q, y_{q,r}) = m_q^{-1}\left({y_q}/({y_q + y_{q,r}})\right).
	\end{align}
When $N$ is high, $y_q + y_{q,r}$ becomes a good estimator for $N$ and the two previous expressions become similar.

	\begin{figure}
		\begin{tabular}{m{3.2cm}m{3.2cm}m{1.0cm}}
			\includegraphics[width=3.6cm]{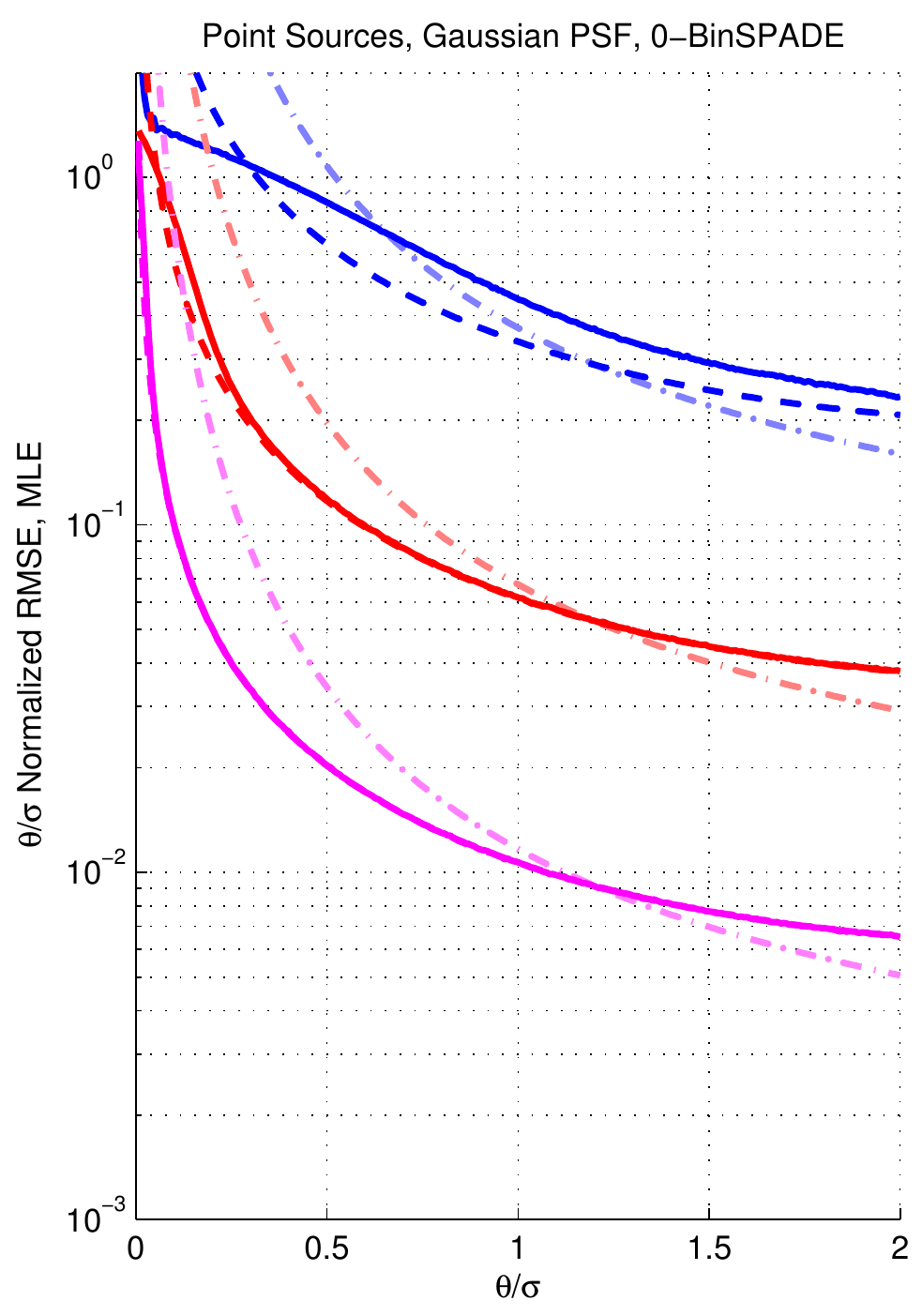} &
			\includegraphics[width=3.6cm]{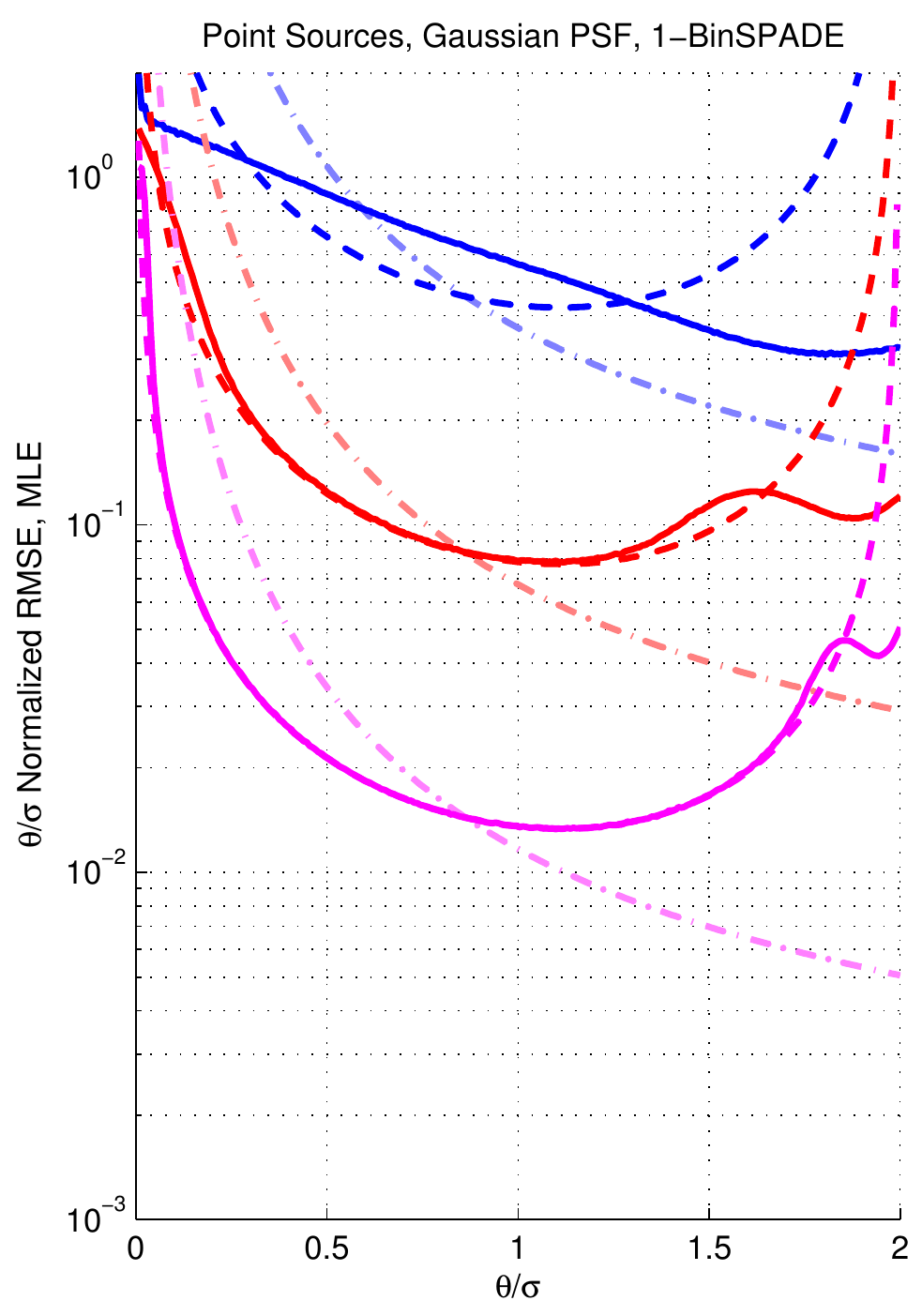} &
			\includegraphics[width=1.5cm]{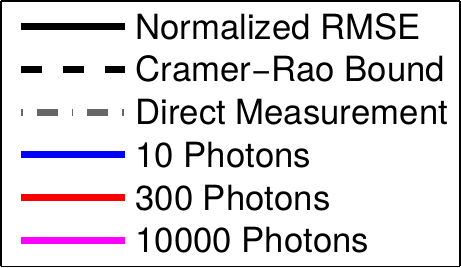}
		\end{tabular}
		\vspace{-3mm}
		\caption{Normalized RMSE plots from Monte Carlo Estimation with more than $65,000$ samples. \textit{(left)} 0-BinSPADE for Gaussian aperture. \textit{(right)} 1-BinSPADE for Gaussian aperture. The estimators for both can be written analytically (with Lambert-W function for the 1-BinSPADE) over the range of $\theta$ shown, where the associated measurement function is invertible.}
		\vspace{-2mm}
		\label{NRMSEPlotGaussian}
	\end{figure}

	\begin{figure}
		\begin{tabular}{m{3.2cm}m{3.2cm}m{1.0cm}}
			\includegraphics[width=3.6cm]{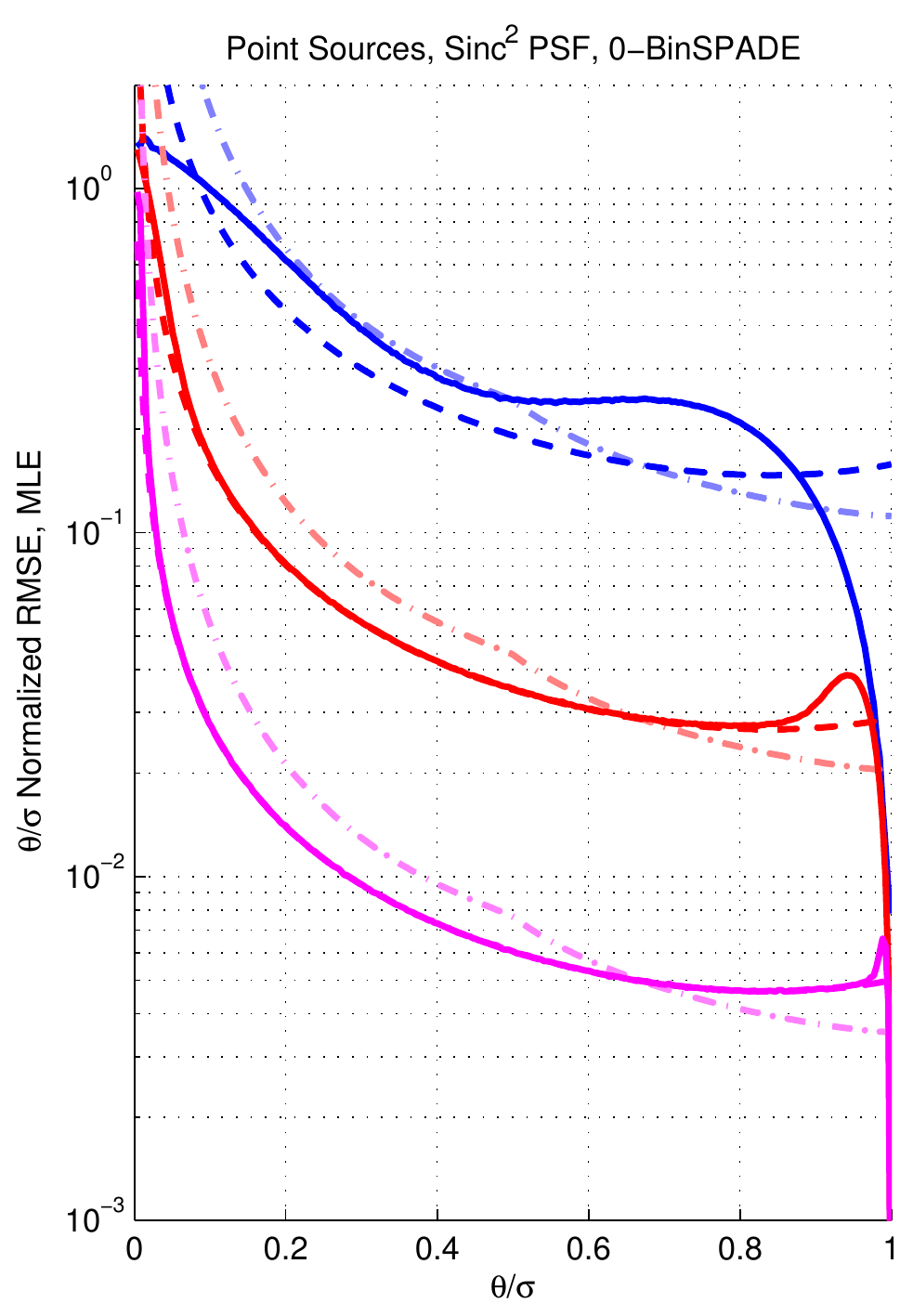} &
			\includegraphics[width=3.6cm]{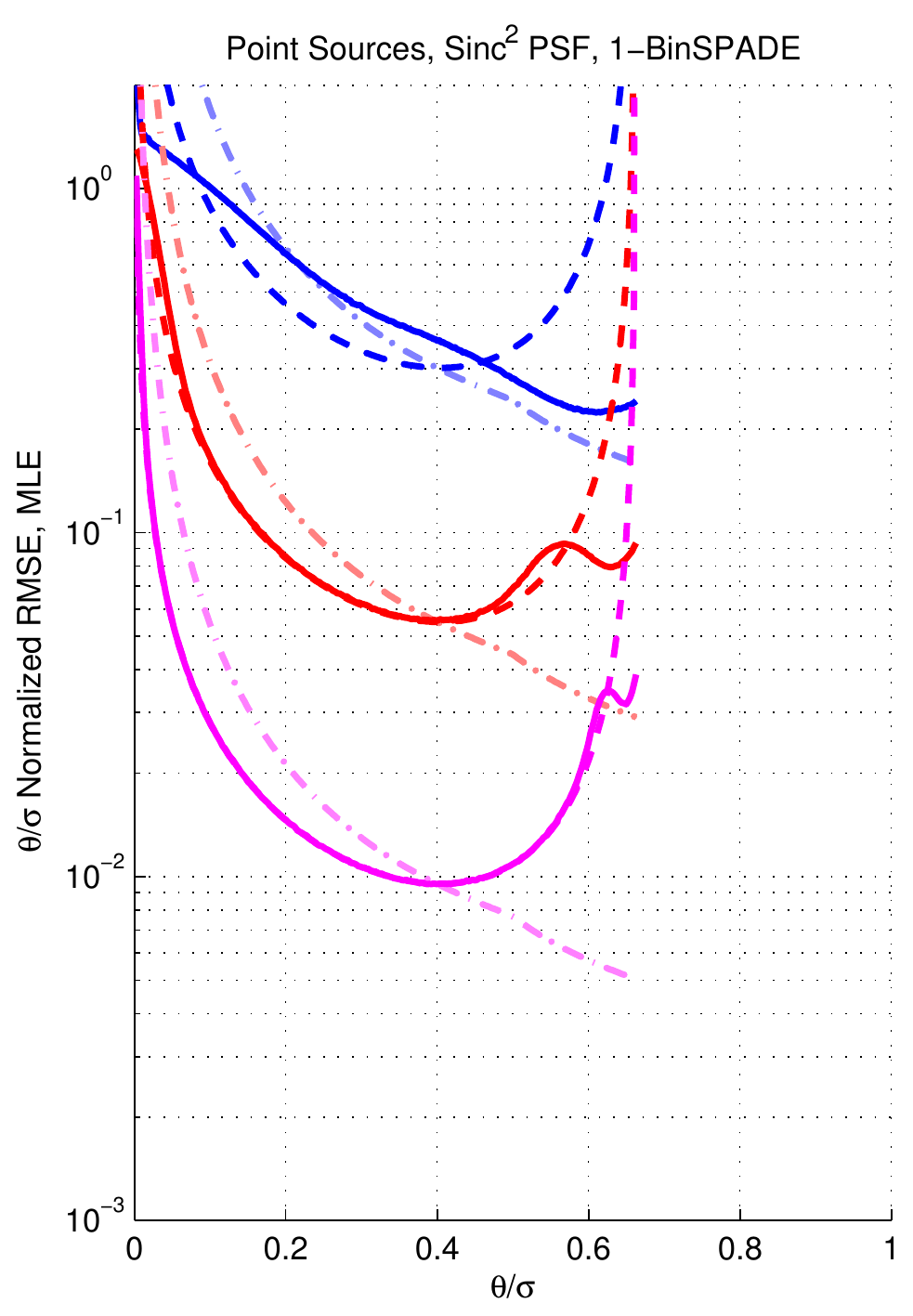} &
			\includegraphics[width=1.5cm]{./plotsLegend.pdf}
		\end{tabular}
		\vspace{-3mm}
		\caption{Normalized RMSE plots from Monte Carlo Estimation. \textit{(left)} 0-BinSPADE for rectangular aperture. \textit{(right)} 1-BinSPADE for hard aperture. Note that the $\theta/\sigma$ range restriction for 1-BinSPADE corresponds to the presence of the first zero in the Fisher Information.}
		\vspace{-2mm}
		\label{NRMSEPlotSinc}
	\end{figure}

	In Figs.~\ref{NRMSEPlotGaussian} and \ref{NRMSEPlotSinc}, we show the performance of MLE for the 0 and 1-BinSPADE receivers for the Gaussian and hard apertures, respectively, for three choices of the total number of detected photons. The MLE performs very close to the CRB at high SNR and generally follows the same trend across all $\theta$. As noted by Tsang~\etal in \cite{MTsang161}, the estimator can become super efficient when it is biased, which happens at very low and large $\theta$. In both graphs, we also plot the numerically-evaluated CRB for image-plane direct measurement using~\eqref{directFisherInformation}. The MSE attained by the BinSPADEs are seen to outperform image-plane detection by a clear margin for $\theta$ much smaller than the Rayleigh separation ($\theta/\sigma=1$). Despite the superior Fisher information of the BinSPADE receivers, the NRMSE highlights the unavoidable precision decay at low $\theta$. Both the BinSPADEs and direct-measurement receivers can reach arbitrary precision in estimating $\theta$ provided they can collect a large enough photon flux (from increased exposure time). But the BinSPADEs remain more efficient than the direct measurement strategy in the sub Rayleigh regime.

\section{Efficient Binary SPADE For Arbitrary ASF}

Consider a general (complex-valued) ASF, $A(x/\sigma)/\sqrt{\sigma}$, $\int_{-\infty}^{\infty} |A(x)|^2\ dx = 1$, in $\mathcal{C}^\infty$, and its autocorrelation function,
	\begin{align}
		\Gamma_A(x') = \int_{-\infty}^{+\infty} \overline{A(x)} A(x+x')\ dx.
	\end{align}

Using~\eqref{twoPointsMeasurementDefinition} the measurement functions associated with measuring the $A(x)$ mode and its orthogonal complement are:
	\vspace{-2mm}
	\begin{align}
		m_{A,0}(\theta) = \left|\Gamma_A\left(\frac{\theta}{\sigma}\right)\right|^2,\quad m_{A,0,r}(\theta) &= 1 - m_{A,0}(\theta).
	\end{align}
The Fisher Information for this generalized 0-BinSPADE receiver evaluates to:
	\vspace{-2mm}
	\begin{align}
		\mathcal{I}_{\text{0-BinSPADE}}(\theta) = \frac{4N}{\sigma^2} \frac{\real{\overline{\Gamma_A^{(1)}\left(\frac{\theta}{\sigma}\right)} \Gamma_A\left(\frac{\theta}{\sigma}\right)}^2}{\left|\Gamma_A\left(\frac{\theta}{\sigma}\right)\right|^2 \left(1-\left|\Gamma_A\left(\frac{\theta}{\sigma}\right)\right|^2\right)}, \label{eqFIAutoCorrelationResidual}
	\end{align}
	where $\Gamma_A^{(1)}$ is the first derivative of $\Gamma_A$. As for the two cases (Gaussian and hard aperture) studied before, this generalized 0-BinSPADE receiver collects all the image-plane energy in the $0$ (ASF) mode for $\theta \to 0^+$. This follows from the peak property of the autocorrelation function and the normalization of the ASF, which gives us $m_{A,0}(0)=1$. It is interesting to note that $\mathcal{I}_{\text{0-BinSPADE}}(\theta)$ is independent of any phase present in the aperture function and is solely based on its intensity profile. Thus, most optical aberrations such as defocusing, spherical aberration or coma, among others, if taken into account into the projection mode, do not degrade the information, unlike the image-plane direct measurement.

Assuming $\Gamma_A(x)$ admits a second order expansion near $\theta = 0$, one can write a Taylor series as follows:
	\begin{align}
		\Gamma_A(x) \underset{\theta\rightarrow 0}{=} 1 + i\beta x - \frac{\alpha}{2} x^2 + \mathcal{O}(x^3), \label{eqAutoCorrExpansion}
	\end{align}
with $\alpha\ge0$, and $\beta\in\mathbb{R}$. One can further show that:
	\begin{align}
		\mathcal{I}_{\text{0-BinSPADE}}(\theta) \underset{\theta\rightarrow 0}{\rightarrow} \frac{4 N}{\sigma^2} (\alpha-\beta^2),\label{eq:I_0_generalASF}
	\end{align}
which is a $\theta$-independent constant as before. Note that a real ASF benefits from the fact that $\beta=0$. Also, from the Wiener-Khinchin theorem, increasing $\alpha$ is equivalent to greater spatial variations in the ASF profile, i.e., sharper or numerous edges. It is simple to verify that~\eqref{eq:I_0_generalASF} reduces to the corresponding formulas for the QFI for the Gaussian ($\alpha=1/4$, $\beta=0$) and hard rectangular apertures ($\alpha=\pi^2/3$, $\beta=0$), attained by the respective $0$-BinSPADE receivers as discussed above.
 
Let us now construct an orthonormal basis for a general ASF by using derivatives of $A(x)$. Let us first note the following identity relating the $q^{\text{th}}$ derivative of $A(x)$ to that of $\Gamma_A$:
	\begin{align}
		\int_{-\infty}^{+\infty} \overline{A^{(q)}(x)} A(x+x')\ dx = (-1)^q \Gamma_A^{(q)}(x')
	\end{align}
Clearly, the functions $(A^{(q)})_q$ need not be orthogonal, and hence cannot be used for parallel measurements. We do a Gram-Schmidt orthogonalization by selecting weights $\omega_{k,q}\in\mathbb{C}$ such that each measurement mode $M_q(x)$ can be written as a linear combination of the $(A^{(q)})_q$ functions,
	\begin{align}
		\frac{1}{\sqrt{\sigma}} M_q\left(\frac{x}{\sigma}\right) = \sum_{k=0}^q (-1)^k\frac{\omega_{k,q}}{\sigma^{q+\frac{1}{2}}} A^{(q)}\left(\frac{x}{\sigma}\right), \label{defGeneralizedMode}
	\end{align}
such that $\left\{M_q\right\}$ forms an orthonormal basis (after removing some possibly identically null functions). It is easy to verify that $\omega_{0,0} = 1$ (due to the energy normalization property of $A(x)$) and hence the first mode ($M_0$) simply equals the ASF. The measurement functions for a simultaneous measurement of the $M_q$ modes can be expressed as:
	\begin{align}
		m_{A,q}(\theta) = \frac{1}{2}\left|\sum_{k=0}^q \frac{\overline{\omega_{k,q}}}{\sigma^k} \Gamma_A^{(k)}\left(\frac{\theta}{\sigma}\right)\right|^2 + \frac{1}{2}\left|\sum_{k=0}^q \frac{\overline{\omega_{k,q}}}{\sigma^k} \Gamma_A^{(k)}\left(-\frac{\theta}{\sigma}\right)\right|^2
	\end{align}

We consider a generalized $1$-BinSPADE receiver associated with the $M_1$ mode. The corresponding measurement function is given by:
	\vspace{-3mm}
	\begin{align}
		m_{A,1}(\theta) &= \frac{\left|\Gamma_A^{(1)}\left(\frac{\theta}{\sigma}\right) - \Gamma_A^{(1)}(0) \Gamma_A\left(\frac{\theta}{\sigma}\right)\right|^2}{-\Gamma_A^{(2)}(0) - \left|\Gamma_A^{(1)}(0)\right|^2},
	\end{align}
and the associated Fisher Information is given by:
	\begin{align}
		\mathcal{I}_{\text{1-BinSPADE}}(\theta) &= \frac{4N}{\sigma^2} \frac{\real{\overline{f^{(1)}\left(\frac{\theta}{\sigma}\right)} f\left(\frac{\theta}{\sigma}\right)}^2}{\left|f\left(\frac{\theta}{\sigma}\right)\right|^2 \left(-f^{(1)}(0) - \left|f\left(\frac{\theta}{\sigma}\right)\right|^2\right)}, \\
		\mbox{with: } f\left(\frac{\theta}{\sigma}\right) &= \Gamma_A^{(1)}\left(\frac{\theta}{\sigma}\right) - \Gamma_A^{(1)}(0) \Gamma_A\left(\frac{\theta}{\sigma}\right).
	\end{align}

Assuming $\Gamma_A(x)$ can be expanded as in (\ref{eqAutoCorrExpansion}), we find that the generalized $1$-BinSPADE receiver attains the same Fisher Information as that of the generalized $0$-BinSPADE receiver: $4N/\sigma^2 (\alpha-\beta^2)$, at $\theta \to 0^+$. As discussed before in the context of Gaussian and hard rectangular apertures, a similar robustness advantage (for 1 over 0-BinSPADE) exists in the general case. Whether or not $\sum_{q=0}^\infty {\mathcal I}_q(\theta) = 4N/\sigma^2 (\alpha-\beta^2), \forall \theta$ holds for a general complex-valued $A(x)$ with our construction of the measurement modes, is left open. In future work, it will also be interesting to investigate a systematic generalization of the optimal modes for a pre-detection mode-sorting based receiver for more complex imaging problems, and to prove that the quantum Fisher Information limit for any incoherent-light imaging problem can be attained by an appropriate receiver that applies a pre-detection linear mode transformation.

	\newpage
	\section*{Appendix}

	\begin{proof}[Fisher Information for a Poisson corrupted process]
		Let $f$ be a function $\mathbb{R} \rightarrow \mathbb{R}^+$ modeling the output of a process on a variable $\theta$ that is corrupted by Poisson noise:
		\begin{align}
			p_f(y|\theta) = \frac{f(\theta)^y \exp(-f(\theta))}{y!},
		\end{align}
		such that $\ln(p_f(y|\theta))$ is twice differentiable with respect to $\theta$. We note $f'$ and $f''$, respectively its first and second derivatives. Then we can write the corresponding Fisher Information as:
		\begin{align}
			\mathcal{I}(\theta) &= -\emean_y\left[\left.\frac{\partial^2}{\partial \theta^2} \ln(p_f(y|\theta)) \right| \theta\right] \\
			&= -\frac{\emean_y[y|\theta]}{f(\theta)} \left(f''(\theta) - \frac{f'(\theta)^2}{f(\theta)}\right) + f''(\theta) \\
			&= \frac{f'(\theta)^2}{f(\theta)}
		\end{align}
	\end{proof}

	\begin{proof}[Fisher Information For Multiple Independent Measurements]
		If we consider the output of multiple independent measurement functions $(f_0(\theta), \ldots, f_Q(\theta))$, all twice differentiable with respect to $\theta$, and their respective outputs $y_0,\ldots,y_Q$, we have for the Fisher Information:
		\begin{align}
			\mathcal{I}(\theta) &= -\emean_{y_0,\ldots,y_Q} \left[\left.\frac{\partial^2}{\partial \theta^2} \ln\left( \prod_{q=0}^Q p_q(y_q|\theta)\right) \right| \theta\right] \\
			&= \sum_{q=0}^Q \frac{f_q'(\theta)^2}{f_q(\theta)} = \sum_{q=0}^Q \mathcal{I}_q(\theta)
		\end{align}
	\end{proof}

	\begin{proof}[Uniform Convergence Of Series Based On Spherical Bessel Functions Of The First Kind]
		We consider the two following series of functions over a finite interval $I$ of $\mathbb{R}$ containing zero, for some fixed positive integer $b$ :
		\begin{align}
			\forall Q\in\mathbb{N},\ x \in I,\ A_{b,Q}(x) &= \sum_{q=0}^Q q^b \frac{\partial}{\partial x} j_q(x)^2 \label{defSeriesTypeA} \\ 
			B_{b,Q}(x) &= \sum_{q=0}^Q q^b \frac{\partial}{\partial x} j_q(x) j_{q+1}(x) \label{defSeriesTypeB} 
		\end{align}
		We have the following loose upper bound on the Spherical Bessel Functions, from their series definition:
		\begin{align}
			\forall q\in\mathbb{N},\ x\in\mathbb{R},\ j_q(x) &= \frac{\sqrt{\pi}}{2^{q+1}} x^q \sum_{k=0}^{\infty} \frac{(-1)^k}{k!\Gamma\left(k+q+\frac{3}{2}\right)}\left(\frac{x}{2}\right)^{2k} \\
			|j_q(x)| &\le \frac{\sqrt{\pi}}{2 q!} \left(\frac{|x|}{2}\right)^q \exp\left(\frac{x^2}{4}\right)
		\end{align}

		We can pick $X$ such that, $\forall x\in I,\ |x|\le X$ and we can test that, for the first series (\ref{defSeriesTypeA}) we have the convergence bound:
		\begin{align}
			&\forall P,Q\in\mathbb{N}, Q>P,\quad \left|A_{b,Q}(x) - A_{b,P}(x)\right| \nonumber \\
			&= \left|\sum_{q=P+1}^Q 2 \frac{q^{b+1}}{x} j_q(x)^2 - 2 q^b j_q(x) j_{q+1}(x)\right| \\
			&\le \pi (1+X^2) \exp\left(\frac{X^2}{2}\right) \sum_{q=P+1}^\infty \frac{q^{b+1}}{2^q q!} X^{2q-1}.
		\end{align}

		As the underlying series is positive, monotonic and converges (by the ratio test), we can choose a $P$ large enough to reduce the remainder, and subsequently the bound, to any small $\epsilon$ we wish. In other words:
		\begin{align}
			\forall \epsilon>0, \exists P\in\mathbb{N}, \forall p,q\ge P, x\in I, \left|A_{b,p}(x) - A_{b,q}(x)\right| < \epsilon .
		\end{align}
		We can thus conclude that the function series is uniformly Cauchy and converges uniformly over $I$.

		For the second series (\ref{defSeriesTypeB}) we have:
		\begin{align}
			&\forall P,Q\in\mathbb{N}, Q>P,\quad \left|B_{b,Q}(x) - B_{b,P}(x)\right| \nonumber \\
			&= \left|\sum_{q=P+1}^Q q^b j_q(x)^2 - \frac{2}{x} q^b j_q(x) j_{q+1}(x) - q^b j_{q+1}(x)^2 \right| \\
			&\le \pi \left(1+\frac{X^2}{2}\right) \exp\left(\frac{X^2}{2}\right) \sum_{q=P+1}^\infty \frac{q^b}{2^q q!} X^{2q},
		\end{align}
		where the same conclusion applies.

		Finally, we consider the new set of series of functions, related respectively to $A_{b,Q}$ and $B_{b,Q}$:
		\begin{align}
			\forall Q,b\in\mathbb{N},\forall x \in I,\ C_{b,Q}(x) &= \sum_{q=0}^Q q^b j_q(x)^2 \label{defSeriesTypeC} \\
			D_{b,Q}(x) &= \sum_{q=0}^Q q^b j_q(x) j_{q+1}(x). \label{defSeriesTypeD}
		\end{align}

		We note that their point-wise convergence can easily be established at $x=0$, as $\forall q>0,\ j_q(0) = 0,\ j_0(x) = 1$ and, considering the previous results, we can apply the Differentiation Theorem to obtain their respective uniform convergence as well as the relations:
		\begin{align}
			C_{b,Q}(x) &= \frac{\partial}{\partial x} A_{b,Q}(x),\,{\text{and}} \label{eqHomogeneous} \\
			D_{b,Q}(x) &= \frac{\partial}{\partial x} B_{b,Q}(x). \label{eqHeterogeneous}
		\end{align}
	\end{proof}

	\begin{proof}[Series of Spherical Bessel Function Of The First Kind]
		We will use the following two results from \cite{Abramowitz64} (Equations 1.10.50 and 1.10.52), where $\Si(x)$ denotes the Sine Integral.
		\begin{align}
			\sum_{q=0}^{\infty} j_q(x)^2 &= \frac{\Si(2x)}{2x},\, {\text{and}} \label{eqAbr01} \\
			\sum_{q=0}^{\infty} (1+2q) j_q(x)^2 &= 1 \label{eqAbr02}
		\end{align}
		By combining them, we obtain:
		\begin{align}
			\sum_{q=0}^\infty q j_q(x)^2 &= \frac{1}{2}\left(1-\frac{\Si(2x)}{2x}\right) \label{eqS1} \\
			\sum_{q=0}^\infty j_{q+1}(x)^2 &= \frac{\Si(2x)}{2x} - j_0(x)^2 \label{eqS2} \\
			\sum_{q=0}^\infty q j_{q+1}(x)^2 &= \frac{1}{2} - \frac{3\Si(2x)}{4x} + j_0(x)^2 \label{eqS3} \\
			\sum_{q=0}^\infty (1+2q) j_{q+1}(x)^2 &= 1 - \frac{\Si(2x)}{x} + j_0(x)^2 \label{eqS4}
		\end{align}

		Thanks to the Differentiation results of the previous lemma (\ref{eqHomogeneous}), we can write from the derivative of (\ref{eqAbr01}):
		\begin{align}
			\sum_{q=0}^\infty j_q(x) j_{q+1}(x) &= \frac{1}{2x}\left(1-j_0(2x)\right). \label{eqS5}
		\end{align}
		
		We then find the following as the solution of a first order linear differential equation involving (\ref{eqS1}) and (\ref{eqS3}) as well as the property (\ref{eqHeterogeneous}) and the constraint that the series is equal to 0 at $x=0$: 
		\begin{align}
			\sum_{q=0}^\infty q j_q(x) j_{q+1}(x) = \frac{\Si(2x)}{4} + \frac{3\sin(2x)}{16x^2} - \frac{1}{2x} + \frac{\cos(2x)}{8x} \label{eqS6}
		\end{align}

		With this result, we can compute the derivative of (\ref{eqS1}) to express the following series and its respective shifted version:
		\begin{align}
			\sum_{q=0}^\infty q^2 j_q(x)^2 &= \frac{1}{8}\Si(2x)\left(\frac{1}{x} + 2x\right) + \frac{\cos(2x)}{8} \nonumber \\
			&\qquad + \frac{j_0(2x)}{8} - \frac{1}{2} \label{eqS8} \\
			\sum_{q=0}^\infty q^2 j_{q+1}(x)^2 &= \frac{1}{8}\Si(2x)\left(\frac{9}{x} + 2x\right) + \frac{\cos(2x)}{8} \nonumber \\
			&\qquad + \frac{j_0(2x)}{8} - j_0(x)^2 - \frac{3}{2} \label{eqS9}
		\end{align}

		The next series is also found as the solution of another differential equation involving (\ref{eqS8}) and (\ref{eqS9}), with the same value constraint at $x=0$ than previously:
		\begin{align}
			\sum_{q=0}^\infty q^2 j_q(x) j_{q+1}(x) &= - \frac{\Si(2x)}{2} - \frac{\sin(2x)}{8x^2} + \frac{x}{3} \nonumber \\
			&\qquad + \frac{1}{2x} - \frac{\cos(2x)}{4x} \label{eqS10}
		\end{align}

		And after deriving (\ref{eqS8}) we obtain:
		\begin{align}
			\sum_{q=0}^\infty q^3 j_q(x)^2 &= -\frac{6x^2+1}{16x} \Si(2x) - \frac{3\sin(2x)+6x\cos(2x)}{32x} \nonumber \\
			&\qquad\qquad + \frac{x^2}{3} + \frac{1}{2} \label{eqS11}
		\end{align}

		Finally, we can combine the results (\ref{eqS3}) to (\ref{eqS11}) to get:
		\begin{align}
			&\sum_{q=0}^\infty (1+2q)\frac{q^2}{x^2} j_q(x)^2 - 2(1+2q)\frac{q}{x}j_q(x)j_{q+1}(x) \nonumber \\
			&\qquad + (1+2q)j_{q+1}(x)^2 = \frac{1}{3}. \label{eS12}
		\end{align}
	\end{proof}

	\begin{proof}[ASF Insensitivity Property]
		Considering any continuous and differentiable ASF $A(x)$ as well as mode $g(x)$, the measurement function $m_g$ for two point sources separated by $2\theta$ can be written as:
		\begin{align}
			m_g(\theta) &= \frac{1}{2\sigma^2} \left|\int_{-\infty}^{+\infty} \overline{g\left(\frac{x}{\sigma}\right)} A\left(\frac{x+\theta}{\sigma}\right)\ dx\right|^2 \nonumber \\
			&\qquad + \frac{1}{2\sigma^2} \left|\int_{-\infty}^{+\infty} \overline{g\left(\frac{x}{\sigma}\right)} A\left(\frac{x-\theta}{\sigma}\right)\ dx\right|^2.
		\end{align}

		We have for the limit of its first derivative:
		\begin{align}
			&\quad \lim_{\theta\rightarrow0} \frac{\partial}{\partial \theta} m_g(\theta) = \lim_{\theta\rightarrow0} \nonumber \\
			&\quad \frac{1}{\sigma^3} \real{\int_{-\infty}^{+\infty} \overline{g\left(\frac{x}{\sigma}\right)} A'\left(\frac{x}{\sigma}\right)\ dx \overline{\int_{-\infty}^{+\infty} \overline{g\left(\frac{x}{\sigma}\right)} A\left(\frac{x}{\sigma}\right)\ dx}} \nonumber \\
			& - \frac{1}{\sigma^3} \real{\int_{-\infty}^{+\infty} \overline{g\left(\frac{x}{\sigma}\right)} A'\left(\frac{x}{\sigma}\right)\ dx \overline{\int_{-\infty}^{+\infty} \overline{g\left(\frac{x}{\sigma}\right)} A\left(\frac{x}{\sigma}\right)\ dx}} \nonumber \\
			&\qquad = 0.
		\end{align}

		We can proceed similarly for the Fisher Information in the case of direct detection. We recall the expression:
		\begin{align}
			\mathcal{I}_{\text{Direct}}(\theta) = \int_{-\infty}^{\infty} \frac{I'(x)^2}{I(x)} dx, \mbox{  with }\, I'(x,\theta) = \frac{\partial I(x,\theta)}{\partial \theta}
		\end{align}

		In the case of two point sources, the normalized intensity profile can be written with the previous ASF as : $I(x)=(|A(x+\theta)|^2+|A(x-\theta)|^2)/2$. One can write for its first derivative:
		\begin{align}
			&\forall x\in\mathbb{R},\ \lim_{\theta\rightarrow 0} I'(x,\theta) = \nonumber \\
			&\lim_{\theta\rightarrow 0} \real{\overline{A'(x+\theta)} A(x+\theta) - \overline{A'(x-\theta)} A(x-\theta)} \nonumber \\
			&= 0 \\
			&\mbox{And : } \lim_{\theta\rightarrow 0} \ \mathcal{I}_{\text{Direct}}(\theta) = 0.
		\end{align}
	\end{proof}

	\begin{proof}[Fisher Information Inequality On Aggregated Measurements]
		Let $m_{S}$ be an aggregated measurement over a collection $S$ of measurement functions from orthogonal modes, \textit{i.e.} $m_S(\theta) =\sum_{q\in S} m_q(\theta)$, all corrupted by Poisson noise. At a location $\theta$ where $\forall q,\ m_q(\theta) > 0$, we have the following measurement function and Fisher Information:
		\begin{align}
			\mathcal{I}_S(\theta) = \frac{\left(\sum_{q\in S} m'_q(\theta)\right)^2}{\sum_{q\in S} m_q(\theta)}
		\end{align}
		We can simplify the notations for the current location into sets $(m_q)$ and $(m'_q)$ and observe that:
		\begin{align}
			&\quad \left(\sum_{q\in S} m_q\right) \left(\sum_{q\in S} \frac{m_q'^2}{m_q}\right) - \left(\sum_{q\in S} m_q'\right)^2 \nonumber \\
			&= \sum_{q\in S} \sum_{r\in S, r>q} \frac{m_r}{m_q} m_q'^2 + \frac{m_q}{m_r} m_r'^2 - 2 m_q' m_r' \label{eqExpandedInequality}
		\end{align}

		Here, if $m'_q$ or $m'_r$ are equal to zero or their product is negative, and with the previous positivity constraint, the sum is clearly positive. Otherwise, one can write:
		\begin{align}
			&\quad \frac{1}{m_q' m_r'}\left(\frac{m_r}{m_q} m_q'^2 + \frac{m_q}{m_r} m_r'^2 - 2 m_q' m_r'\right) \nonumber \\
			&= \frac{m_r}{m_q} \frac{m_q'}{m_r'} + \frac{m_q}{m_r} \frac{m_r'}{m_q'} - 2 \ge 0
		\end{align}

		As we have, $m_r m_q' / m_q m_r' > 0$ and $\forall x>0,\ x+1/x\ge2$. Thus, (\ref{eqExpandedInequality}) is always positive and we can conclude with the inequality:
		\begin{align}
			\frac{\left(\sum_{q\in S} m_q'(\theta)\right)^2}{\sum_{q\in S} m_q(\theta)} = \mathcal{I}_S(\theta) \le \sum_{q\in S} \frac{m_q'(\theta)^2}{m_q(\theta)} = \sum_{q\in S} \mathcal{I}_q(\theta)
		\end{align}
	\end{proof}

	\begin{remark}[Generalized projection modes for a BinSPADE receiver]
		One can expand the generalized mode expression (\ref{defGeneralizedMode}) in the case of the two first modes for a normalized ASF $A(x)$ into :
		\begin{align}
			\frac{1}{\sqrt{\sigma}} M_0\left(\frac{x}{\sigma}\right) &= \frac{1}{\sqrt{\sigma}} A\left(\frac{x}{\sigma}\right) \\
			\frac{1}{\sqrt{\sigma}} M_1\left(\frac{x}{\sigma}\right) &= \frac{1}{\sqrt{\sigma}} \frac{-A^{(1)}\left(\frac{x}{\sigma}\right) + \Gamma_A^{(1)}(0) A\left(\frac{x}{\sigma}\right)}{\sqrt{-\Gamma_A^{(2)}(0) - \left|\Gamma_A^{(1)}(0)\right|^2}},
		\end{align}
		where $A^{(1)}(x)$ is the first derivative of the ASF, $\Gamma_A$ is the autocorrelation of $A$ and $\Gamma_A^{(1)}$, $\Gamma_A^{(2)}$ are respectively its first and second derivatives. Here, one can note, as the autocorrelation is hermitian, that $\Gamma_A^{(1)}(0)$ is purely imaginary. Thus in the case of a purely real ASF it is equal to zero and the first mode can be constructed with only $A^{(1)}(x)$.
	\end{remark}

	\begin{remark}[Fisher Information of a leaky BinSPADE receiver]
		Considering a leakage factor $0<\epsilon<1$, we define the efficiency of a BinSPADE receiver as $\rho = 1-\epsilon$. We have the following expressions of the Fisher Information for 0 and 1-BinSPADE respectively; first, for a Gaussian aperture:
		\begin{align}
			\mathcal{I}_{\text{0-BinSPADE},\rho}(\theta) &= \frac{N\rho}{\sigma^2} \frac{\frac{\theta^2}{4\sigma^2}}{\exp\left(\frac{\theta^2}{4\sigma^2}\right)-\rho} \\
			\mathcal{I}_{\text{1-BinSPADE},\rho}(\theta) &= \frac{N\rho}{\sigma^2} \frac{\left[1 - \frac{\theta^2}{4\sigma^2}\right]^2}{\exp\left(\frac{\theta^2}{4\sigma^2}\right)-\rho\frac{\theta^2}{4\sigma^2}}
		\end{align}

		For a rectangular aperture:
		\begin{align}
			\mathcal{I}_{\text{0-BinSPADE},\rho}(\theta) &= \frac{4N\rho}{\sigma^2} \frac{\left(\frac{\sigma}{\theta}\cos\left(\frac{\pi\theta}{\sigma}\right) - \frac{\sigma^2}{\pi\theta^2} \sin\left(\frac{\pi\theta}{\sigma}\right)\right)^2}{1 - \rho j_0\left(\frac{\pi\theta}{\sigma}\right)^2} \\
			\mathcal{I}_{\text{1-BinSPADE},\rho}(\theta) &= \frac{12 N \rho}{\sigma^2} \frac{\sigma^2}{\theta^2} \frac{\left[\sin\left(\frac{\pi\theta}{\sigma}\right) - 2 j_1\left(\frac{\pi\theta}{\sigma}\right)\right]^2}{1 - 3 \rho j_1\left(\frac{\pi\theta}{\sigma}\right)^2}
		\end{align}

		Finally, for a generic ASF $A(x)$, properly normalized and in $\mathcal{C}^2$:
		\begin{align}
			\mathcal{I}_{\text{0-BinSPADE},\rho}(\theta) = \frac{4N\rho}{\sigma^2} \frac{\real{\overline{\Gamma_A^{(1)}\left(\frac{\theta}{\sigma}\right)} \Gamma_A\left(\frac{\theta}{\sigma}\right)}^2}{\left|\Gamma_A\left(\frac{\theta}{\sigma}\right)\right|^2\left(1-\rho\left|\Gamma_A\left(\frac{\theta}{\sigma}\right)\right|^2\right)} \label{eqLeaky0BinSpace}
		\end{align}

		\begin{align}
			\mathcal{I}_{\text{1-BinSPADE},\rho}(\theta) &= \frac{4N\rho}{\sigma^2} \frac{\real{\overline{f^{(1)}\left(\frac{\theta}{\sigma}\right)} f\left(\frac{\theta}{\sigma}\right)}^2}{\left|f\left(\frac{\theta}{\sigma}\right)\right|^2 \left(-f^{(1)}(0) - \rho \left|f\left(\frac{\theta}{\sigma}\right)\right|^2\right)} \label{eqLeaky1BinSpace} \\
			\mbox{with: } f\left(\frac{\theta}{\sigma}\right) &= \Gamma_A^{(1)}\left(\frac{\theta}{\sigma}\right) - \Gamma_A^{(1)}(0) \Gamma_A\left(\frac    {\theta}{\sigma}\right).
		\end{align}

		One can notice from the common normalization of the ASF $A(x)$ that the autocorrelation $\Gamma_A(\theta)$ is equal to one at $\theta=0$; and as it is also hermitian, that $\Gamma_A^{(1)}(0)$ is purely imaginary. Hence, in (\ref{eqLeaky0BinSpace}) the numerator always tend to zero for $\theta\rightarrow 0$, but the denominator only does so if the efficiency $\rho$ is exactly equal to one. Thus, for any $\rho<1$, the generalized 0-BinSPADE does not collect information for narrow separation angles : $\mathcal{I}_{\text{0-BinSPADE},\rho<1}\rightarrow 0$ near $0$.

		On the other hand, with the expansion (\ref{eqAutoCorrExpansion}) one can develop the limit of (\ref{eqLeaky1BinSpace}) as :
		\begin{align}
			\lim_{\theta\rightarrow 0} \mathcal{I}_{\text{1-BinSPADE},\rho}(\theta) &= \frac{4N\rho}{\sigma^2} (\alpha-\beta^2).
		\end{align}
		Thus, the generalized 1-BinSPADE is resilient to leakage in the deep sub-Rayleigh limit.

	\end{remark}

\end{document}